\documentclass[11pt]{article}

\usepackage[margin=2.5cm]{geometry}
\usepackage{setspace}
\usepackage{subcaption}
\usepackage[utf8]{inputenc}
\usepackage{url}
 
\usepackage{hyperref}
\expandafter\let\csname equation*\endcsname\relax
\expandafter\let\csname endequation*\endcsname\relax
\usepackage{amsmath}
\usepackage{amsfonts}
\usepackage{graphicx}
\usepackage{amsmath,xparse}
\usepackage{amsthm}
\usepackage{xcolor}
\usepackage{color}
\usepackage{rotating}\usepackage{booktabs}
\usepackage{floatrow}
\usepackage{titlesec}
\usepackage{url}

\usepackage[utf8]{inputenc}
\usepackage[english]{babel}
\newtheorem*{theorem*}{Theorem}

\usepackage[utf8]{inputenc}

\usepackage{float}
\usepackage{tikz}
\usetikzlibrary{positioning,arrows}
\tikzset{
	m/.style={circle,draw,fill=gray!20,minimum size=5},outer sep=2pt}

\usepackage{amsmath,amsfonts,amssymb,amsthm, mathrsfs}
\usepackage{csquotes}

\usepackage{dcolumn}

\newcolumntype{d}[1]{D{.}{.}{#1} } 

\usepackage{bbm}

\usepackage{graphicx,wrapfig, tikz, cancel, hyperref, multirow,rotating}
\usepackage{natbib}
\usepackage{enumerate}
\usepackage{pdflscape} 

\theoremstyle{plain}

\newtheorem{prop}{Proposition}
\newtheorem{lemma}{Lemma}[section]
\newtheorem{cor}{Corollary}[section]

\theoremstyle{definition}

\newcommand{\dd}{\mathrm{d}}

\newcommand{\EE}{\mathbb{E}}

\usepackage{authblk}

\title{The social value of overreaction to information
\thanks{This paper was initially part of a different project titled \enquote{Learning, overreaction, and the wisdom of the crowds}. We wish to thank for useful comments Nicola Gennaioli, Philip Matejka, Filippo Massari, Fabrizio Panebianco, Antonio Rosato, Alex Teytelboym, Fernando Vega-Redondo, Giovannni Immordino, Giacomo Battiston, Riccardo Franceschin and participants to the 2019 SMYE, the 2021 ASSET, the 2021 GRASS, and seminar participants in Bocconi and Napoli.}}	

\date{}

\author[a]{Matteo Bizzarri} 
\author[b]{Daniele d'Arienzo}

\affil[a]{University of Naples Federico II and CSEF}
\affil[b]{Capital Fund Management and Nova Business School}

\usepackage{savesym}

\usepackage[final]{changes}

\savesymbol{comment}

\usepackage{comment}

\begin{document}
	
\maketitle

\begin{abstract}

 
We study the welfare effects of overreaction to information in the form of diagnostic expectations in markets with asymmetric information, and the effect of a simple intervention in the form of a tax or a subsidy. A large enough level of overreaction is always welfare-decreasing and can rationalize a tax on financial transactions. A small degree of overreaction to private information can both increase or decrease welfare. This is because there are two competing externalities: an information externality, due to the informational role of prices, and a pecuniary externality, due to the allocative role of prices. When the information externality prevails on the pecuniary externality, the loading on private information in agents' trades is too small compared to the welfare optimum: in this case, a small degree of overreaction is welfare-improving. 

    
\end{abstract}

\bigskip



\pagebreak

\section*{Introduction}

Information aggregation is understood to be one of the fundamental roles of markets, and financial markets in particular. As a consequence, a large literature has studied the welfare properties and the social value of information in markets, from \cite{hayek2013use} to, e.g. \cite{angeletos2007efficient}. In doing so, it is crucial to understand how agents make inferences from the information they receive: for example, traders in financial markets constantly update their beliefs about valuations of financial assets, as a consequence of changes in market prices, fundamentals, and investment choices of other traders. There is growing evidence that agents' updating rules depart from Bayesian rationality in the form of over or underreaction to information, namely agents react in the correct direction to news, however too little or too much in magnitude with respect to the Bayesian benchmark. 
In this paper, we ask: how do such departures of individual updating rules from Bayesian rationality impact welfare and informational efficiency in financial markets? Can a simple intervention, such as a tax or a subsidy, mitigate inefficiencies?


\par
\added{To formalize departures from Bayesian rationality in a parsimonious way, we rely on the memory-based model of \emph{diagnostic expectations}, introduced in \citep{bordalo2018diagnostic}. Diagnostic expectations is a one parameter, forward-looking departure from Bayesian updating: when computing their posterior distribution, diagnostic agents in the model react in the correct direction to news, however too little or too much in magnitude. The model is one of the simplest ways to reconcile anomalies in forecast data \citep{bordalo2020overreaction} and experiments \citep{afrouzi2023overreaction}.
Moreover, it has been used to rationalize several facts about macro-financial variables, such as credit cycles (\citeauthor{bordalo2018diagnostic}, \citeyear{bordalo2018diagnostic}), stock return puzzles (\citeauthor{bouchaud2019sticky}, \citeyear{bouchaud2019sticky} and \citeauthor{bordalo2018diagnostic}, \citeyear{bordalo2018diagnostic}), interest rates (\citeauthor{d2020maturity}, \citeyear{d2020maturity}) and even the likelihood of a financial crisis (\citeauthor{maxted2024macro}, \citeyear{maxted2024macro}). This literature is reviewed in \citep{gennaioli2018crisis}. 
Moreover, the model is portable and has been used to shed light on belief formation beyond financial applications, such as understanding stereotypes (\citeauthor{bordalo2016stereotypes},\citeyear{bordalo2016stereotypes}) and political beliefs (\citeauthor{bonomi2021identity},\citeyear{bonomi2021identity}).}

\replaced{The majority of the papers above find that data are consistent with overreaction to information. However, some papers find that in short time horizons data display underreaction \citep{bouchaud2019sticky}, and more generally that the level of overreaction may depend on the time horizon \citep{d2020maturity}. So, given that ours is an abstract setting, we allow for different parameter values representing both over and underreaction.
}{To formalize departures from Bayesian rationality in a parsimonious way, we follow the logic of the diagnostic expectations model (\citeauthor{bordalo2016stereotypes}, \citeyear{bordalo2016stereotypes} and \citeauthor{bordalo2018diagnostic}, \citeyear{bordalo2018diagnostic}) that formalizes overreaction (and underreaction) in beliefs as a parametric deviation from Bayesian updating. 
Biased agents depart from Bayesian rationality in computing posterior beliefs by under/over-reacting to recent information.}
We embed over-reacting agents in a market game in which agents submit conditional bids, or schedules, that depend on the market price and a private signal. We adopt the tractable linear-quadratic Gaussian setting from \cite{vives2017endogenous}. 

In this environment, there are two sources of information: the private signal and the (public) market price. We adopt the \emph{diagnostic expectations equilibrium} of \cite{bordalo2020diagnostic}, in which prices are formed in equilibrium given agents' trade choices, and agents correctly understand this mechanism, but their posterior expectation about the fundamental value is distorted due to over/under reaction to both private information (the private signal) and public information (the market price). In particular, in our context, the bias does not come from (possibly partially) failing to realize that other traders also understand the information contained in prices, as in the \enquote{cursed equilibrium} of \cite{eyster2019financial} or the \enquote{partial equilibrium thinking} of \cite{bastianello2023partial}.
The main difference between \cite{bordalo2020diagnostic} and our work is that their focus is on bubbles rather than welfare and taxes.\footnote{Moreover, they use a model with CARA utility and inelastic supply, whereas, for tractability, we follow \cite{vives2017endogenous} using a model with elastic supply and quadratic utility.}

In a version of this model with standard Bayesian agents, \cite{vives2017endogenous} highlights two competing externalities: a learning externality, due to the fact that agents do not internalize that their actions reveal information by changing the informativeness of the price as a signal of the underlying value; and a pecuniary externality, due to the fact that agents do not internalize that, conditioning their trade on the price, they also change how the price reacts to the underlying value. As a consequence, the loading on private information can be either too high with respect to the efficient benchmark (if the pecuniary externality prevails) or too low (if the learning externality prevails). Both cases are possible, for different values of the parameters.

We characterize the equilibrium in a tractable linear-quadratic setting.
When agents display overreaction, agents trade more aggressively for the same private signal, because they overweight the information contained in it. As a consequence, they increase the informativeness of the price as a public signal of the value. However, this increase is not sufficient to offset the first-order effect, and so the loading on the private signal in agents' actions is larger than it would be for Bayesian agents. So, overreaction changes the relative importance of the learning and the pecuniary externality with respect to the benchmark model. As a consequence, the price reveals more information than in an economy with Bayesian agents.

Having characterized the equilibrium, we study the effect of overreaction on welfare. The externality that prevails in the Bayesian benchmark determines the sign of the welfare effect a small level of overreaction, and so it can be positive or negative. However, for a large enough level of overreaction, a further increase in the diagnostic bias is always decreasing welfare. Then, we explore whether introducing a small quadratic tax or subsidy can be optimal. We show that when the overreaction parameter is large enough, the introduction of a small tax is always welfare-improving. Such result can offer a rationalization of a Tobin-type tax (\citeauthor{tobin1978proposal}, \citeyear{tobin1978proposal}) on financial transactions, for reasons related to the interaction of a behavioral bias (diagnostic expectations) and informational efficiency, that are distinct both from arguments relating to curbing speculation (as in \citeauthor{stiglitz1989using}, \citeyear{stiglitz1989using} and \citeauthor{summers1989financial}, \citeyear{summers1989financial}), and arguments arising from disagreement in agents' evaluations such as in \cite{davila2023optimal}, and thus can be seen as complementary to such arguments.\footnote{Such a tax has been the subject of a long debate and is still a important issue in economic policy: it has been first advocated by Keynes, is currently in place in multiple countries (such as UK and Sweden), and is the object of a European Commission official proposal since 2011.} When instead agents underreact strongly enough, a small subsidy is optimal.\footnote{A concrete example of a policy that can be compared to a subsidy are the tax incentives for investment in retirement plans and pension funds, present in many countries: for example the tax-deductibility of 401(k) plan contributions in the USA (\citeauthor{engen2000effects}, \citeyear{engen2000effects}); similar policies are present in many countries, such as Italy and the UK (\citeauthor{whitehouse2005taxation}, \citeyear{whitehouse2005taxation}). } 
When overreaction is close to zero, the welfare effect of a tax depends on the balance between the learning and pecuniary externality in the Bayesian benchmark. So the
model implications for the optimality of a tax crucially depend on the degree of agents' overreaction to information.

Our work is related to three literatures: the literature on overreaction and related biases in information processing, the literature studying taxes in the presence of behavioral biases, especially on financial transactions, and the literature on the social value of information. Our contribution is to show how overreaction can be welfare improving via mitigating the learning externality: that, is, overreaction can have a \enquote{social value}. However, when overreaction is large enough, it
can rationalize a tax on financial transactions, even in the presence of the learning externality. The literature on overreaction in finance and macroeconomics has mostly focused on identifying and measuring overreaction and on its explanatory power for rationalizing various macroeconomic phenomena (\citeauthor{bordalo2022overreaction}, \citeyear{bordalo2022overreaction}). Some papers have explored macroeconomic policy under overreaction or exuberance, such as \cite{maxted2024macro}, which also finds a positive welfare effect, that does not work through the learning externality but a balance sheet mechanism.
\cite{davila2023prudential} explore macro-prudential policy implications with extrapolative beliefs. The fact that overreaction helps learning via revealing more information is similar to the effect of overconfidence in the social learning model of \cite{bernardo2001evolution}: they study a simple sequential learning model instead of a financial market and so, in their setting, only the learning externality is present, but not the pecuniary externality.

\added{The literature on behavioral finance has studied models that incorporate related biases in information processing. 
In the cursed equilibrium of \citeauthor{eyster2019financial} and \cite{bayona2022competition} agents neglect the informational content of the price. This can be seen as an extreme form of underreaction to the price signal. Instead, in the diagnostic expectation model we use, agents overreact or underreact to all information in the same way. \cite{eyster2019financial} does 
 not study welfare; while \cite{bayona2022competition} shows that cursedness can improve welfare. \cite{mondria2022costly} study costly information processing, that has similar implications to underreaction to the price, in that agents do not consider adequately the information in the price signal; they show that this can give rise to excess volatility: this is the opposite implication we get from underreaction, because in our case when agents underreact they do so also with respect to their private information. None of these papers focus on the effect of tax/subsidy schemes.}

\added{Another related bias is overconfidence.
The main difference between overreaction and overconfidence is that overconfident agents overestimate the precision of their information, but their updating is still Bayesian, as in: \cite{kyle1997speculation}, \cite{bernardo2001evolution}, \cite{sandroni2007overconfidence}, \cite{daniel2001overconfidence}, \cite{daniel2001overconfidence}. So, overconfidence cannot explain the predictability of forecast errors observed in the data \citep{bordalo2020overreaction}, \citep{afrouzi2023overreaction}. Moreover, even if the posterior is biased in the same direction in both models, the posterior expectation is still a convex combination of the prior and the signals, whereas the expectation of overreacting agents can overshoot and lie outside of such a convex combination. \cite{bordalo2022overreaction} argue that this fact means that overreaction can rationalize facts about the behavior of bubbles that overconfidence cannot. Again, none of these papers focus on the effect of tax/subsidy schemes.
}

While the taxation literature has studied various behavioral biases, for example, related to attention and salience as in \cite{goldin2015optimal}, \cite{moore2021optimal}, \cite{farhi2020optimal},
the literature specifically on taxation \emph{of financial transactions} has mostly focused on rational models: \cite{auerbach2004generalized}, \cite{rochet2023taxing}, \cite{adam2017stock}, \cite{buss2016intended}, at most with heterogeneous priors as in \cite{davila2023optimal}. The literature on the social value of information has also mainly focused on Bayesian agents, e.g. \cite{angeletos2007efficient}, \cite{angeletos2009policy}, \cite{bayona2018social}, \cite{colombo2014information}. An exception is \cite{ostrizek2021acquisition}, that study a strategic setting in which agents follow the cursed equilibrium model of \cite{eyster2005cursed} and \cite{eyster2019financial}, showing that cursedness can improve welfare: their mechanism works through information acquisition and not through the pecuniary externality like ours.

The next section introduces the model, Section \ref{characterization} describes the equilibrium characterization, Section \ref{effect} describes our results, and Section \ref{conclusion} concludes. All the proofs are in the Appendix.

\section{The model}

Our model closely follows \cite{vives2017endogenous}, in its financial market interpretation, except for the behavioral bias due to diagnostic expectations.\footnote{\cite{vives2017endogenous} studies different interpretations of the same abstract model, one being agents in a financial market, and another firms competing in schedules. For our purposes we stick to the interpretation of agents trading in a financial market.} 
We consider a financial market populated by informed speculators and liquidity suppliers. There is only one asset traded. 

\paragraph{Informed agents} There is a continuum of informed speculators indexed by $i \in [0,1]$ and represented with the density $f$. Informed speculators 
face quadratic transaction costs. Each of them can decide her position $D_i$ with respect to the only asset exchanged, where short sales are allowed ($D_i$ can be negative). 

The profit of an informed agent $i$ holding $D_i$ units of the asset when the market price is $p$ is:
\[
u_i(D_i,p,V)= (V-p)D_i-\frac{1}{2}\gamma D_i^2 
\]
where $V$ is the (unobservable) fundamental value of the asset, and the quadratic term represents transaction costs. Equivalently, it can be considered a form of (non constant) risk aversion.\footnote{The quadratic functional form makes the model very tractable. A similar approach is followed in \cite{vives2014possibility}.}
Informed speculators have a prior over the fundamental value $V$ which is Gaussian: $V \sim \mathcal{N}(0,\tau_0^{-1})$. They also have access to a private signal $s_i$ that, conditional on $V$, follows a Gaussian distribution: $s_i\mid V \sim \mathcal{N}\left(V,\tau_{\varepsilon}^{-1}\right)$. Moreover, $s_i$ is independent of $s_j$ for $i \neq j$, conditionally on $V$: $s_i\perp s_j \mid V$. 

In the following, various steps involve the integration of a continuum of random variables over $[0,1]$. We follow the literature\footnote{See \cite{vives2010information}.} \emph{defining} the integral over a continuum of independent random variables $(X_i)_{i\in [0,1]}$ as $\int X_i \dd i:=\int \EE (X_i)\dd i$ whenever the map $\EE (X_i)$ is integrable (that is always the case in our setting). This implies that a form of the Law of Large numbers holds, so that, conditionally on $V$, we have $\int s_i \dd i=V$. This is going to be the only property of such an integral we need.\footnote{The most commonly used approach to formalize the integral over a continuum of random variables is the one of \cite{uhlig1996law}. Since the only property we are going to need is the Law of Large numbers, we avoid these technical issues and directly assume it.} 
We denote the total demand from all informed agents as $\overline{D}=\int D_i\dd i$.

\paragraph{Diagnostic expectations}
Agents update their prior using the private signal $s_i$ and also the information contained in the price $p$ but, crucially, not in a Bayesian way.
If the price depends on the fundamental $V$ and the noise $S$ according to $p=A+BV-CS$, then $(p-A)/B$ is a Gaussian random variable of mean $V$ and precision $B^2/C^2\tau_S$: the agents understand this dependence and use it for their  updating.
So, after observing private signal $s_i$ and the price $p$, the Bayesian posterior distribution of the belief on the fundamental $V$ is a Normal with parameters:
\begin{align*}
 \EE(V\mid s_i,p)&=\frac{\tau_{\varepsilon}}{\tau_{\varepsilon}+\tau_0+B^2/C^2\tau_S}  s_i+\frac{B^2/C^2\tau_S}{\tau_{\varepsilon}+\tau_0+B^2/C^2\tau_S}\frac{p-A}{B}\\
 Var(V\mid s_i,p)&=\left(\tau_0+\tau_{\varepsilon}+B^2/C^2\tau_S\right)^{-1}
\end{align*}

Our informed agents do \emph{not} hold these beliefs because we assume that they over / underreact to information according to the diagnostic expectations model of \cite{bordalo2018diagnostic} and \cite{bordalo2020diagnostic}. Namely, their posterior beliefs follow a Gaussian with the same variance, but expectation equal to:
\begin{equation}
\EE^{\theta,i}(V\mid s_i,p):= \EE(V\mid s_i,p)+\theta( \EE(V\mid s_i,p)- \EE(V))
\label{diagnostic}
\end{equation}
where $\theta \in (-1,+\infty)$ represents the strength of the diagnostic bias.
When $\theta>0$ agents \emph{over-react} to the information: when the information leads them to revise their prior expectation upwards ($\EE(V\mid s_i,p)>\EE(V)$), they revise it upwards more than a Bayesian would: $\EE^{\theta,i}(V\mid s_i,p)> \EE(V\mid s_i,p)$; while if the information leads to a downward revision ($\EE(V\mid s_i,p)<\EE(V)$), they revise it downwards more than a Bayesian would: $\EE^{\theta,i}(V\mid s_i,p)< \EE(V\mid s_i,p)$. The case of Bayesian agents corresponds to $\theta=0$. For $\theta<0$ we instead obtain \emph{under}-reaction: agents revise their priors \emph{less} than a Bayesian would; for $\theta\to -1$, agents do not revise their prior at all. We allow for both overreaction and underreaction, since both have been found to be consistent with the data (even if underreaction only with a very short time horizon, \citeauthor{bouchaud2019sticky}, \citeyear{bouchaud2019sticky}).

\added{The model is a parsimonious characterization of \cite{kahneman1972subjective} \enquote{representativeness heuristic}. When forming posterior beliefs, agents overweight representative traits of the group, which are those traits objectively more likely in that group relative to a benchmark group. One popular example of this heuristic is to assess the likelihood that a person is red-haired (trait) given the information that the person is Irish.
  Such probability is exaggerated because being red-haired is \emph{representative} of Irish people, i.e. is more likely among Irish relative to the rest of the world (benchmark group). The representativeness bias is consistent with biased beliefs in seemingly unrelated domains, from stereotypes (\citeauthor{bordalo2016stereotypes}, \citeyear{bordalo2016stereotypes}), to race (\citeauthor{arnold2018racial}, \citeyear{arnold2018racial}). A deeper foundation of this heuristic is rooted in the functioning of human memory: representative traits came to mind more often, which generates the bias (\citeauthor{bordalo2023memory}, \citeyear{bordalo2023memory}).
\cite{bordalo2016stereotypes} formalize the heuristic by assuming that, when estimating the probability of a trait $s$ (e.g., being red-haired), after observing a piece of information $G$ (the person is Irish), the conditional probability density $f(s|G)$ is inflated/deflated by (an increasing function of) the likelihood ratio $\frac{f(s|G)}{f(s |G_0)}$, where $G_0$ is a reference group (the general population, rather than the Irish population). 
A popular specification (because of its tractability with Gaussian, exponential and power law distributions) is to assume that the distorted posterior density $f^{\theta}(s\mid G)$ is, up to a normalization constant, equal to:
\begin{align}
    f^{\theta}(s\mid G)\propto f(s\mid G) \left(\frac{f(s|G)}{f(s |G_0)}\right)^{\theta}
    \label{representativeness} 
\end{align}
The parameter $\theta$ modulates the strength of the effect. The case of Bayesian agents corresponds to $\theta=0$. For $\theta>0$ when $f(s|G)>f(s |G_0)$ agents overestimate $f(s|G)$ (the probability that an Irish person has red hair).
For $\theta\to \infty$ agents completely neglect the prior.
For $\theta<0$ we instead obtain \emph{under}-reaction: agents revise their priors \emph{less} than a Bayesian would; for $\theta\to -1$, agents do not revise their prior at all. So, it is common to consider the meaningful range of $\theta$ as $(-1,\infty)$.}

\added{In the setting of financial markets or more generally in information updating, the idea is that a positive signal is more representative of a good underlying fundamental than a signal equal to the average of the prior. So, agents displaying the representativeness bias, when trying to assess the posterior distribution of the asset value $V$ after observing information $G=(s_i,p)$, have a posterior that follows Equation \eqref{representativeness}, where $s$ is equal to the fundamental value $V$, and
where the benchmark $G_0$ is a pair of signals that are exactly confirming the prior expectation: $G_0: \,(s_i=\EE(V),p=(\EE(V)-A)/B)$.
So, traders displaying the representativeness bias for $\theta>0$ overestimate the likelihood of a good state when observing a good signal, or underestimate it when $\theta<0$. These are what are known as \emph{ diagnostic expectations}. \cite{bordalo2020diagnostic} shows that when $f$ follows a Gaussian distribution, such as the one we use, the diagnostic expectation bias yields the formula \eqref{diagnostic} for the updating.}

\added{The model can be microfounded based on friction in memory retrievals (\citeauthor{bordalo2023memory},\citeyear{bordalo2023memory}), costly information processing \citep{afrouzi2023overreaction}, or rational inattention \citep{gabaix2019behavioral}. The details depend on the specific case, but a general idea is that failure to properly take into consideration all the past information can generate overreaction to the most recent information.
}

\paragraph{Liquidity suppliers} \added{As in \cite{vives2017endogenous}, liquidity suppliers have an elastic supply function. In particular, they trade according to the aggregate (inverse) supply function $p=-\mu_S-S+\beta \overline{D}$. $S$ is a random variable
distributed as $S \sim \mathcal{N}(0,\tau_S^{-1})$, representing the noise in the demand. The parameter $\mu_S$ is a constant that we can think of as a shifter of the random variable $S$, that we include for generality but has little effect on the efficiency properties. Instead, the slope of the supply $\beta>0$ is going to be important, because it regulates how prices react to quantities and the strength of both the learning and the pecuniary externality. 
Classic noise traders, as in \cite{grossman1980impossibility} are a special case of this specification in which $\beta \to \infty$, $\tau_S\to \infty$ and $\tau_S\beta^2 =\tau_S'>0$. In this case, the aggregate supply is independent of prices, and simply a random variable with precision $\tau_S'$.} In the welfare measure \eqref{welfare_benchmark}, we include the surplus of the liquidity suppliers, defined as is standard as the area below the supply curve: $\int_0^{\overline{D}} p(q)\dd q$.\footnote{If we were to exclude the liquidity traders from welfare calculations, there would still be a scope for intervention, as even in the Bayesian benchmark \cite{vives2017endogenous} shows that the learning and pecuniary externality would still be present, even if the precise expression would change.} An alternative interpretation that does not rely on the concept of noise traders (and so might have a clearer welfare interpretation) is that there is an entrepreneur that can issue equity yielding a dividend $V$, with a preference for retaining shares (control) of the firm measured by $S$. This is explored formally in the Appendix \ref{liquidity}.

\paragraph{Equilibrium}

Agents compete choosing demand schedules, that is, functions $D_i$, that map values of the private signal $s_i$ and the price $p$ into real numbers $D_i(s_i,p)$ representing the net demand of agent $i$. 

We follow (\cite{bordalo2020diagnostic}) in looking for a \emph{diagnostic expectations equilibrium}, that is analogous to the Bayesian Nash equilibrium of the game in schedules of \cite{vives2017endogenous}, except that agents are not Bayesians but have diagnostic expectations. Namely, we look for a set of demand schedules $D_i$ and a pricing function $P$ that satisfy:

\begin{enumerate}

\item Individual optimization: the demand function $D_i$ maximizes the (diagnostic) expected utility of the trader $i$ given the observation of the private signal $s_i$ and the price $p$, formally: $D_i(s_i,p) \in \arg\max_{x_i} \{\EE^{\theta,i} [u_i(x_i,p,V)\mid s_i,P(S,V)=p]\}$;

\item Market clearing: the pricing function clears the market, that is, the relation $P(S,V)=-\mu_S-S+\beta\int D_i(P(S,V),s_i) 
\dd i$ holds for any realization of $S$, $V$, and each $s_i$. 

\end{enumerate}

Similar to \cite{vives2017endogenous}, \replaced{we restrict attention to linear equilibria, namely equilibria where the function $g$ is linear.}{the functional forms assumptions guarantee that the equilibrium pricing function is linear} So, determining the equilibrium reduces to finding the coefficients $A$, $B$ and $C$ such that $P(S,V)=A+BV-CS$ satisfies the conditions above.

\paragraph{The welfare measure}

We follow \cite{vives2017endogenous} in considering our welfare measure the total surplus, defined as informed trader surplus plus the surplus of the liquidity suppliers:
\begin{equation}
W=\EE\left(\left(\mu_S+S-\beta\frac{1}{2}\overline{D}\right)\overline{D}+\int \left(V D_i -\frac{\gamma}{2} D_i^2\right) 
\dd i\right)
\label{welfare_benchmark}
\end{equation}


In the alternative interpretation of the asset supply as arising from an entrepreneur issuing equity, this expression represents the surplus of the informed traders plus the profit of the entrepreneur, that is also equivalent to the utilitarian welfare in this economy.
Note that the expectations that appear in the expression are all taken from the perspective of Bayesian agents. In doing this, we interpret the agents' deviation from the Bayesian benchmark as a proper \enquote{mistake}, not as a taste or preference feature, following a standard approach in the behavioral economics literature, e.g.: \cite{o2006optimal}, \cite{spinnewijn2015unemployed}, and the survey by \cite{mullainathan2012reduced}.\footnote{There is another, more conceptual reason. To compute the ex-ante welfare from the perspective of a diagnostic decision maker would require to specify how the decision maker predicts her future behavior \emph{once she receives the information}: is she aware of her bias or not? this would require considerably more assumptions than simply compute the welfare from the perspective of a Bayesian agent, so we follow the latter approach.
} 


In this context the first best allocation, that would realize if agents could pool their information, would be the complete information allocation, since by the law of large numbers $\int s_i \dd i=V$. It is convenient to study welfare in terms of \emph{welfare loss} from such an allocation.
The first best allocation solves:
\[
\max_{D_i} W= \left(\mu_S+S-\beta\frac{1}{2}\overline{D}\right)\overline{D}+\int \left(V D_i -\frac{\gamma}{2} D_i^2\right) \dd i
\]
and since the agents are ex-ante identical is a symmetric allocation, that we denote $D^o$. Denote $W^o$ the aggregate welfare in such an allocation. Define the welfare loss of some allocation $(D_i)_{i\in[0,1]}$ from the first best as $WL=W^o-W$, where $W$ is the welfare in the allocation $(D_i)_{i\in[0,1]}$. The following lemma from \cite{vives2017endogenous} characterizes the welfare loss from the first best:
\begin{lemma}
	
At the allocation $(D_i)_{i \in [0,1]}$ the welfare loss from the first best allocation $D^o$ is
	\[
	WL=\EE(W^*-W)=(\beta+\gamma)\frac{1}{2}\EE(\overline{D}-D
^o)^2+\frac{\gamma}{2}\EE \int (D_i-\overline{D})^2\dd i
	\]

 \label{welfare_loss_vives}

\end{lemma}

The interpretation of the above expression is that the welfare loss results from two parts, that \cite{angeletos2007efficient} name, respectively, \enquote{variance} and  \enquote{dispersion}: the first relative to the departure of the aggregate demand from its first best level, the second relative to the cross-sectional dispersion of trades across agents. The effect of information (and thus overreaction to information) results from this trade-off: precise information means a small aggregate deviation from the first best, but a large dispersion, because precise information means traders trade more aggressively. The welfare impact of overreaction will result from this fundamental trade-off.

\section{Equilibrium characterization}
\label{characterization}

In this section we illustrate the equilibrium and the welfare benchmark.

The equilibrium strategies are linear, and have the functional form:
\[
D_i=\alpha s_i+\eta \EE(V\mid p)-\eta_pp
\]
where $\alpha$ is the \emph{loading on private information}, $\eta$ is the \emph{loading on public information}, and $\eta_p$ is the \emph{loading on the price}. In the following, it is going to be useful to distinguish the effect on welfare of the loading on public information $\eta$ and on private information $\alpha$. So, in the Proposition below first we solve for the equilibrium assuming agents use a strategy of the above form, for generic loadings, to show the effect of the loadings on the precision of public information.
Then we compute the loadings in the diagnostic expectations equilibrium.

\begin{prop}

If agents choose the demand function:
\[
D_i=\alpha s_i+\eta \EE(V\mid p)-\eta_pp
\]
then the pricing function is linear: $ P(S,V)=A+BV-CS$, and the precision of public  information is: $\tau=\tau_0+\alpha^2\beta^2\tau_S$.

In particular, in the diagnostic expectations equilibrium, we have:
\begin{align*}
\alpha &=a(\theta+1)\\    
\eta &=(\theta+1)(1-\gamma a)\\
\eta_p&=\dfrac{1}{\gamma}\\
\end{align*}
so that $\tau=\tau_0+a^2(\theta+1)^2\beta^2\tau_S$, where $a$ is the unique solution of the equation:
	\begin{equation}
	\gamma a=\frac{\tau_{\varepsilon}}{\tau_{\varepsilon}+\tau }
 \label{loading}
	\end{equation}

The equilibrium pricing function satisfies:
  \[
 P(S,V)=A+BV-CS
	\]
for the coefficients:
\[
A=\frac{-\gamma \mu_S}{\gamma+\beta}
\quad
B=\beta\frac{(\theta+1)}{\gamma+\beta}
\quad 
C=\frac{1}{(\gamma+\beta) a}
\]
 
Moreover:
 \[
 \EE(V\mid p)=\frac{\beta^2a^2\tau_S(\theta+1)^2}{\tau_0+\beta^2a^2\tau_S(\theta+1)^2}(p-A)/B
 \]

\label{decentralized_solution1}	
	
\end{prop}

\subsection{Properties of the equilibrium with diagnostic expectations}

We collect some of the positive properties of the equilibrium in the next Corollary.

\begin{cor}
\begin{enumerate}

In equilibrium the following properties hold:

\item The sensitivity to private information $\alpha$ is increasing in $\theta$;

\item The precision of the price as a signal of the value $B^2/C^2\tau_S$ is increasing in $\theta$;

\item The volatility of the price $Var(p)$ is increasing in $\theta$.
    
\end{enumerate}
\label{positive}

\end{cor}

Point 1 yields the fundamental mechanism of what follows: overreaction increases the sensitivity to private information. This is immediate by construction when fixing the precision of the public signal but, in equilibrium, overreaction also affects such precision, because more information is revealed. This indirect effect on the precision of the price, though, is not strong enough to counteract the main effect, and so the loading $\alpha$ increases in $\theta$.

Point 2 shows that since with overreaction the sensitivity to private information is higher, the price reacts more to the true value than it would in the Bayesian case, and so the precision of the price as a signal of the value is higher: this is analogous to what happens in the model of \cite{bordalo2020diagnostic}.

Point 3 shows that the price displays \emph{excess volatility} under overreaction. This is because overreaction induces agents to trade more aggressively, so this generates larger price movements. Excess volatility of financial markets is a well known empirical regularity, and this result shows that also in our setting can be rationalized by overreaction to information, as in \cite{bordalo2023long}, \cite{bordalo2022overreaction}.


\section{The effect of overreaction}
\label{effect}

In this section we study the effect of overreaction. First, as a benchmark, we illustrate the welfare analysis of the Bayesian model with $\theta=0$.

\subsection{The Bayesian benchmark}

Define $a^*$ as the loading on the private signal at the market solution in the Bayesian benchmark: that is the solution of equation \eqref{loading} for $\theta=0$.
Define $a^T$ as the solution of:
\[
a^T=\frac{\tau_{\varepsilon}}{\gamma (\tau(a^T)+\tau_{\varepsilon})+\beta \tau(a^T)-\Delta(a^T)}
\]
where $\Delta(a^T)=\frac{(1-\gamma a^T)^2\beta^2\tau_S\tau_{\varepsilon}}{\gamma\tau(a^T)}$.
\cite{vives2017endogenous} shows that the market solution is second-best efficient if and only if $a^*=a^T$. In particular, using the fact that in the market solution when the loading on private info is $a$ then the loading on public information is $\eta=(1-\gamma a)/\gamma$, we can think to the welfare loss as a function of $a$ and we have: $\frac{\dd WL}{\dd a}>0 \iff a^*>a^T$. In particular, the loading on private information at the market equilibrium $a^*$ can either be too high or too low from a welfare perspective.
 This is because of the interplay between a \emph{learning externality} and a \emph{pecuniary externality}. The learning externality derives from the informational role of the price and is well understood: agents decisions to trade reveal information to other agents through the price, but agents do not internalize this effect in the market equilibrium. This force pushes the sensitivity $a^*$ to be too low with respect to the second best. The pecuniary externality derives from the allocative role of the price, and derives from the fact that agents decisions affect how the price correlates to the true value $V$, but do not internalize this in the market equilibrium. This externality pushes the sensitivity $a^*$ to be \emph{too large}. Summing up:

\begin{enumerate}

    \item if $a^T<a^*$, this means that the learning externality is stronger;

    \item if $a^T>a^*$, this means that the pecuniary externality is stronger.

\item if $a^*=a^T$ the two externalities exactly balance each other and the market equilibrium maximizes welfare.

\end{enumerate}

\subsection{Overreaction and the information loadings}

The reason why in the Bayesian case it is sufficient to look at the loading on private information is that in the second-best (team) solution the loading on \emph{public} information has the same relation with the loading on private information as in the decentralized solution: $\eta=\dfrac{1}{\gamma}-a^T$. As a consequence, the loading on public information is at the second-best level if and only if the loading on private information is at the second-best level; and when
the loading on private information is higher than the efficient level the loading on public information is too low and vice versa. This breaks down with diagnostic expectations: it is possible that both loadings are too high or too low with respect to the efficient benchmark.
The next Proposition characterizes this behavior.

\begin{prop}

\begin{enumerate}
    \item 
There is a unique value $\theta'$ such that the loading on private information is at the efficient level: $\alpha(\theta')=a^T$. Moreover, $\theta'>0$ if and only if in the Bayesian benchmark $a^*<a^T$.

\item There is a unique value $\theta''$ such that the loading on public information is at the efficient level: $\eta(\theta'')=\dfrac{1}{\gamma}-a^T$. Moreover, $\theta''>0$ if and only if in the Bayesian benchmark $a^*>a^T$.

\item The two values are the same, $\theta'=\theta''$, if and only if agents are Bayesians: $\theta'=\theta''=0$, and the Bayesian benchmark is efficient: $a^*=a^T$.
\end{enumerate} 

\label{bayesian}
    
\end{prop}

So, the key way in which over/underreaction affects welfare is to change the equilibrium loadings on information. If in the Bayesian benchmark the learning externality is stronger, so that $a^*<a^T$, then a sufficiently strong level of overreaction is always sufficient to reproduce the efficient loading on private information. When the pecuniary externality is stronger ($a^*>a^T$), a sufficiently high level of underreaction can reproduce the efficient loading on private information. The analogous happens for the loading on public information but, crucially, part 3) clarifies that no distortion $\theta$ can reproduce the efficient level for both. The Proposition clarifies the key trade-off of an increase in overreaction: the welfare effect depends on the balance of the effect on the two loadings.

\subsection{Welfare decomposition}

The endogenous loadings on private  and public information $\alpha$ and $\eta$ are crucial to understand the efficiency properties of the equilibrium.
In the following Lemma, we provide a decomposition of the welfare loss 
that is going to be useful in the following.

\begin{lemma}
	
	In equilibrium, we can decompose the welfare loss \eqref{welfare_benchmark} as $WL= WL^B+WL^D$:
	\begin{align}
	WL^B&=\frac{1}{2}\frac{(1-\gamma \alpha)^2}{(\beta+\gamma)}\frac{1}{\tau}+\frac{\gamma \alpha^2}{2\tau_{\varepsilon}}
 \\
	WL^{D}&=\dfrac{(1-\gamma \alpha-\gamma \eta)^2}{2(\beta+\gamma)}\left(\frac{1}{\tau_0}-\frac{1}{\tau}\right)   
	\end{align}
	
where $1-\gamma \alpha-\gamma \eta=\theta$.	
	
	\label{decomposition}
\end{lemma}

The first term $WL^B$ is the welfare loss that would realize for Bayesian agents having loading on private information equal to $\alpha$. The second term $WL^D$ represents the additional bias that diagnostic expectations add \emph{beyond} the change in $\alpha$. \added{It represents the welfare loss due to the inefficient relation between the loading on private information and the loading on public information. In the Bayesian benchmark and team solution $\eta=\dfrac{1}{\gamma}-\alpha$, so the term $WL^D$ vanishes. Instead, with diagnostic expectations, we have $1-\gamma \alpha-\gamma \eta=\theta$. In particular, this term comes from the fact that the weight of public information will overshoot or undershoot with respect to the optimal value, depending on whether $\theta>0$ or $\theta<0$.
} This is useful to separate the direct effect of overreaction from the effect on the loading $\alpha$.

\subsection{Welfare effect}

The following proposition characterizes the effect of overreaction on welfare.

\begin{prop}

In $\theta=0$ we have that the welfare loss is increasing if and only if the pecuniary externality prevails in the Bayesian benchmark. That is, formally:
\[
\frac{\dd WL}{\dd \theta}>0\mid_{\theta=0}>0 \iff a^*>a^T
\]

Moreover, there are thresholds $\theta^*$, $\theta_*$ such that for $\theta>\theta^*$ we have $\frac{\dd WL}{\dd \theta}>0$, and for $\theta<\theta_*$ we have $\frac{\dd WL}{\dd \theta}<0$.

\label{welfare}
\end{prop}

The proposition shows that, when overreaction is close to zero, its welfare impact depend solely on the balance of externalities in the Bayesian case: in particular, if $a^*<a^T$, so that the learning externality prevails, overreaction is welfare improving. The key mechanism driving the result is that overreaction increases the sensitivity to private information $\alpha=a(\theta+1)$, \added{and also increases the sensitivity to public  information $\eta$ (as Proposition \ref{bayesian} describes):}
\[
\frac{\dd WL}{\dd \theta}=\frac{\partial WL}{\partial \alpha}\frac{\dd \alpha}{\dd \theta}+\frac{\partial WL^D}{\partial \eta}\frac{\dd \eta}{\dd \theta}
\] 
The increase of $\alpha$ has the effect of making the price more sensitive to the true value, that has two implications: first, this makes the price a better signal of the value, mitigating the information externality; second, it exhacerbates the pecuniary externality. The increase in $\eta$, instead, has only the effect of increasing the term related to the over/undershooting of expectations $WL^D$.
From Lemma \ref{decomposition} we can conclude that the loading on public information affects only $WL^D$, and indeed the term $WL^D$ is minimized for $\eta=1-\gamma \alpha$, that is true only when $\theta=0$.\footnote{This can also be 
 seen from the fact that the term $WL^D$ is second order in $\theta$.} This is because the precision of public information is only affected by the loading $\alpha$, not $\eta$. As a consequence, $\frac{\partial WL^D}{\partial \eta}\mid_{\theta=0}=0$ and also $\frac{\partial WL^D}{\partial \alpha}\mid_{\theta=0}=0$, and, since $\frac{\dd \alpha}{\dd \theta}>0$ by Corollary \ref{positive}, we have:
 \[
sgn\left(\frac{\dd WL}{\dd \theta}\right)=sgn\left(\frac{\partial WL}{\partial \alpha}\frac{\dd \alpha}{\dd \theta}+\frac{\partial WL^D}{\partial \eta}\frac{\dd \alpha}{\dd \theta}\right)=sgn\left(\frac{\partial WL^B}{\partial \alpha}\right)
\]
\added{The sign of the welfare impact is given by the sign of $\frac{\partial WL}{\partial \alpha}$, that is positive if and only if the pecuniary externality is stronger at the Bayesian benchmark, from Proposition \ref{bayesian}.}
  
When the overreaction parameter is far from $0$, the term $WL^D$ instead becomes important. Such a term incorporates the expected mistake the agents make overestimating (underestimating) $V$ when they get positive (negative) information. The second part of the Proposition says that if the overreaction parameter $\theta$, and the consequent expected error, has magnitude large enough, positive or negative, then moving further from the Bayesian benchmark can only reduce welfare. To sum up: a limited amount of overreaction can have a positive effect, depending on the interplay of prediction error, information externality and pecuniary externality.

\begin{figure}[t]
    \centering

   \begin{subfigure}[t]{0.45\textwidth}
    \includegraphics[width=\textwidth]{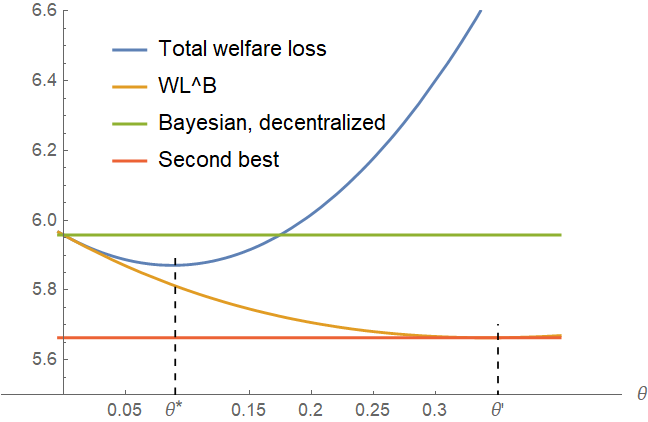}
    \caption{The parameters are: $\gamma=3$, $\beta=0.1$, $\tau_0=\tau_{\varepsilon}=0.01$, $\tau_{S}=50$. The loadings satisfy $a^*=0.079<a^T=1.12$, and the thresholds $\theta^*=\theta_*=0.09<\theta'=0.35$.}
    \label{Fig:overreaction}       
   \end{subfigure} %
   \begin{subfigure}[t]{0.45\textwidth}
    \includegraphics[width=\textwidth]{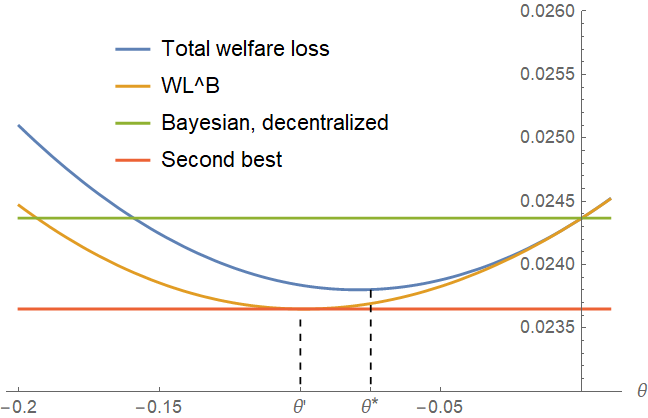}
    \caption{The parameters are: $\gamma=3$, $\beta=2$, $\tau_0=\tau_S=1$, $\tau_{\varepsilon}=5$. The loadings satisfy $a^*=1.64>a^T=1.54$, and the thresholds $\theta^*=\theta_*=-0.075>\theta'=-0.1$.
    }
    \label{Fig:underreaction}       
   \end{subfigure} 

\caption{A graphical representation of the welfare loss as a function of $\theta$ for different values of the parameters. }

    \label{Fig:overunder} 
\end{figure}

\added{
In Figure \ref{Fig:overunder}, we can see a graphical representation of the welfare loss as a function of $\theta$ for different values of the parameters. The blue line represents the total welfare loss $WL$, the orange line represents $WL^B$, the horizontal green line the welfare loss at the decentralized market equilibrium for $\theta=0$, and the red line the welfare loss at the second best (team) solution. The optimal value of $\theta$ is denoted $\theta^*$. The minimum of the welfare loss $WL^B$ is reached for the value $\theta'$ of Proposition \ref{bayesian}, such that $a^*(\theta')=a^T$. In the left example, we have that $a^*<a^T$: the learning externality is stronger, so in the Bayesian benchmark the loading on private information is too small. As Proposition \ref{welfare} finds, the graph shows that around $\theta=0$ a small increase in overreaction decreases the welfare loss. The mechanism works through overreaction increasing the loading on private information. However, 
since also the loading on public information increases, we have that the value of $\theta'$, where the loading on private information is at the efficient level, is too large: at that level the welfare loss is increasing again.
Indeed, the optimal value of overreaction is reached for a value $\theta^*$ smaller than $\theta'$. In both the above figures, we find $\theta^*=\theta_*$. Analogously, in the right figure, $a^*>a^T$, so the pecuniary externality prevails, and indeed a small decrease in $\theta$ improves welfare. 
}

\subsection{Policy}

We have seen that in this economy there are multiple inefficiencies due to the fact that agents might trade too much or too little relative to what would be the optimum given their private signals. These inefficiencies are already present
 in the Bayesian case: moreover, the diagnostic bias can exhacerbate (or not) these inefficiencies. Since the inefficiencies stem from the departures of the amounts traded from the second best, now we explore whether a tax (or subsidy) on quantities exchanged can be used to correct the inefficiencies and provide higher welfare. \cite{vives2017endogenous} shows that, in the Bayesian case, a quadratic tax/subsidy can implement the second-best level of the loading on private information $a^T$. In this section we ask a related question, that is: when does the introduction of a small tax improves welfare, and when a small subsidy instead?

A linear tax/subsidy here cannot improve welfare: it would simply shift uniformly all the demands, but would leave the loading on private and public information unaffected: so it would simply add an additional term $t^2$ to the welfare loss, contributing to the volatility term: so the introduction of such a linear tax/subsidy would never be optimal. The natural next step is to explore a quadratic tax/subsidy $\delta$.

Formally, we assume that when agents trade a volume $|D_i|$, they have to pay an additional amount $\frac{1}{2}\delta D_i^2$, where if $\delta<0$ this is understood to be a subsidy. Both buyers and sellers have to pay the tax. So, the payoff of the informed speculators becomes:
\[
u_i= (V-p)D_i-\frac{1}{2}(\gamma+\delta) D_i^2 
\]
We assume $\delta>-\gamma$ so that the problem of the agents remains concave.
Since the tax is levied also on the liquidity suppliers, the inverse demand becomes: $p=-\mu_S-S+(\beta+\delta)\overline{D}$. In the Appendix we show that the results are qualitatively the same if the tax is levied on informed speculators only.

We follow the assumption in \cite{vives2017endogenous} that the revenues/payments from this tax/subsidy are rebated in a lump-sum amount $T$, to satisfy budget balance. So, the rebate $T$ does not affect the optimal choice of the agents. In the model with a tax/subsidy, to obtain the demand of agent $i$ we simply have to substitute $\gamma+\delta$ to $\gamma$ and $\beta+\delta$ to $\beta$ in the equations of Proposition \ref{decentralized_solution1}, to obtain the following expressions for the equilibrium loadings:
\begin{align}
    \alpha(\delta)&=a(\delta)   (\theta+1)\nonumber\\
    \eta(\delta)&=\left(\dfrac{1}{\gamma+\delta}-a(\delta)\right)(\theta+1)=\dfrac{\theta+1}{\gamma+\delta}-\alpha(\delta)\nonumber \\
    \eta_p(\delta)&=\dfrac{1}{\gamma+\delta}
    \label{demand_tax}
\end{align}
where $a(\delta)$ solves:
\[
(\gamma+\delta) a(\delta)=\frac{\tau_{\varepsilon}}{\tau_{\varepsilon}+\tau(a(\delta))}
\]
and $\tau(a(\delta))=a(\delta)^2(\theta+1)^2(\beta+\delta)^2\tau_S$. The effect of the tax is to reduce the incentive to trade: this means that a higher tax affects \emph{both} the loading on private information and the loading on public information. In the following the loadings are always functions of $\delta$, so we suppress the functional dependence to lighten notation.

The total amount paid from the informed speculators is $\frac{\delta}{2} \int D_i^2\dd i$ and the one paid by the entrepreneur is $\frac{\delta}{2}\overline{D}^2$, and the total revenues collected must equal the rebate, so: $T=\frac{\delta}{2} \int D_i^2\dd i+\frac{\delta}{2}\overline{D}^2$.
So the welfare loss with respect to the first best 
 is:
\[
W^*-\left(W-\frac{\delta}{2}\overline{D}^2 -\frac{\delta}{2} \int D_i^2\dd i+T\right)=W^*-W
\]
because the additional terms cancel out thanks to budget balance.
So we conclude that the welfare loss satisfies the same expression as in Lemma \ref{welfare_loss_vives}. 

\added{The expression \eqref{demand_tax} shows that the tax affects both the loading on private information, and the loading on the price. The next Proposition characterizes the effect of the tax on the loadings.} 

\begin{prop}

In the diagnostic expectation equilibrium of the model with the tax/subsidy $\delta$, we have:

\begin{enumerate}

    \item The loading on private information $\alpha$ is decreasing in the tax: $\dfrac{\dd \alpha}{\dd \delta}<0$. Moreover, there always exist a unique $\delta^*$ such that $\alpha(\delta^*)=a^T$.

\item The loading on public information $\eta$ can be both increasing  or decreasing in $\delta$.

\item The loading on the price $\eta_p$ is decreasing in $\delta$: $\dfrac{\dd \eta_p}{\dd \delta}<0$.

\item The equilibrium is second-best efficient if and only if $\theta=0$, $a^*=a^T$ and $\delta=0$. 

\end{enumerate}

\label{loading_tax}
    
\end{prop}

\added{
The tax tends to decrease the loadings, because it tends to decrease trade. Indeed, the loading on private information $\alpha$ is decreasing in the tax. However, the tax has an ambiguous effect on the informativeness of the price: $B^2/C^2=\alpha^2(\beta+\delta)^2$, because it increases the slope of the demand $\beta+\delta$. So, since it can increase the precision of public information, it has an ambiguous effect on the loading on public information $\eta$.}

\added{Point 1) shows that it is always possible to find a tax level $\delta^*$ that implements the second-best level of the loading on private information, meaning that $\alpha(\delta^*)=a^T$.
However, since the tax distorts all the loadings, including the price loading, there is no tax level that can achieve second-best efficiency. This is easiest to see noting that the second-best efficient loading on the price is equal to $1/\gamma$, so the only tax that can achieve it is $\delta^*=0$, even in the Bayesian case $\theta=0$, since the price loading does not depend on $\theta$. Then point 4 follows from the case with no tax studied in Proposition \ref{bayesian}. 
}

If we cannot achieve the second best, can we at least improve welfare with a tax/subsidy? The next Proposition answers affirmatively. \added{It shows the expression of the welfare loss in the equilibrium with the tax, shows that there is always a finite optimal level of tax/subsidy, and} studies the welfare effect of the introduction of a \emph{small tax}, formally characterized as the derivative of the welfare loss, computed in $\delta=0$: $\frac{\dd WL}{\dd \delta}\mid_{\delta=0}$. When $\frac{\dd WL}{\dd \delta}\mid_{\delta=0}<0$ a small positive tax decreases the welfare loss, and so we say that \emph{a small tax is welfare improving}. When the opposite is true, we say that \emph{a small subsidy is welfare improving}.  

\begin{prop}

The welfare loss from the introduction of a tax $\delta$ is:
\begin{align*}
WL^{\delta}&=\frac{(1-(\gamma+\delta) \alpha)}{\left(\beta+\gamma+2\delta\right)^2\tau}\left((1-(\gamma+\delta) \alpha)+\frac{4\delta}{\beta+\gamma}(1+(\beta+\delta)\alpha)\right)+\frac{4\delta^2(\mu_S^2+\tau_S^{-1}+\tau_0^{-1})}{(\beta+\gamma)^2(\beta+\gamma+2\delta)^2}\\
&+\left(\frac{1}{\beta+\gamma+2\delta}\right)^2\left(\theta^2-\frac{4\delta\theta}{\beta+\gamma}\left(1-\frac{\tau_0}{(\beta+\delta)\alpha \tau_S}\right)\right)\left(\frac{1}{\tau_0}-\frac{1}{\tau}\right)+\frac{\gamma \alpha^2}{\tau_{\varepsilon}}
\end{align*}

Moreover:
\begin{enumerate}


\item If $\theta$ is large enough \added{(overreaction strong enough)}, the introduction of a small tax is welfare improving: $\frac{\dd WL^{\delta}}{\dd \delta}\mid_{\delta=0}<0$;

\item If $\theta$ is small enough \added{(underreaction strong enough)}, the introduction of a small subsidy is welfare improving: $\frac{\dd WL^{\delta}}{\dd \delta}\mid_{\delta=0}>0$;

\item If $\theta=0$, a tax could be either welfare improving or decreasing depending on the parameters. For $a^*=a^T$ a small tax is welfare decreasing \added{if and only if $\alpha 
   \beta  \tau _S (\alpha 
   (\beta +\gamma )-1)+\tau
   _0>0$}. 

\end{enumerate}

\label{prop_tax}
\end{prop}

A tax $\delta$ decreases the total amount traded, and in so doing it also changes  the loadings: an increase in $\delta$ decreases $\alpha$, $\eta$ and $\eta_p$. The expression of the welfare loss above sums up these direct and indirect effects.
When $\theta$ is large enough we obtain $\frac{\dd WL}{\dd \delta}\mid_{\delta=0}>0$. This is because when $\theta$ goes to infinity, then also $\alpha$ and $\eta$ do. So, the amount traded is larger than at the efficient level, and a tax partially corrects this distortion, so it is welfare improving. When $\theta$ is small enough the reasoning is analogous, obtaining a subsidy instead of a tax.

When $\delta=0$ the indirect effect $\frac{\partial WL^{\delta}}{\partial \alpha}$ is the same as without the tax: so it is positive or negative according to whether $a^*>a^T$ or vice versa.
\replaced{However, here this is not the only first order effect. Here the effect of the tax here works not only through the demand of the informed traders (and their loadings), but also through the slope of supply of the liquidity suppliers $\beta+\delta$. Moving $\delta$ away from zero here has two effects: first, it distorts (downward) the amount traded, creating an average discrepancy between the first best and the equilibrium; second, it affects both the strength of the learning externality via the precision of public information ($\tau=\tau_0+\alpha^2(\beta+\delta)^2\tau_S$), and the strength of the pecuniary externality, because it directly changes how the price reacts to quantity. 
 So, even when $a^*=a^T$, the tax/subsidy can be welfare improving depending on the interplay of these effects. Indeed, in the next paragraph, where we explore the case of a tax that affects only the informed traders, these additional effects are absent and for $\delta=0$ the welfare effect of the tax is solely determined by whether $a^*>a^T$ or vice versa.
 }{But, contrary to Proposition \ref{welfare}, the direct effect of the tax is first order here, so it is not sufficient to look at the balance of the learning and pecuniary externality alone to understand the sign. }

\begin{figure}[t]
    \centering
    \includegraphics[width=0.7\textwidth]{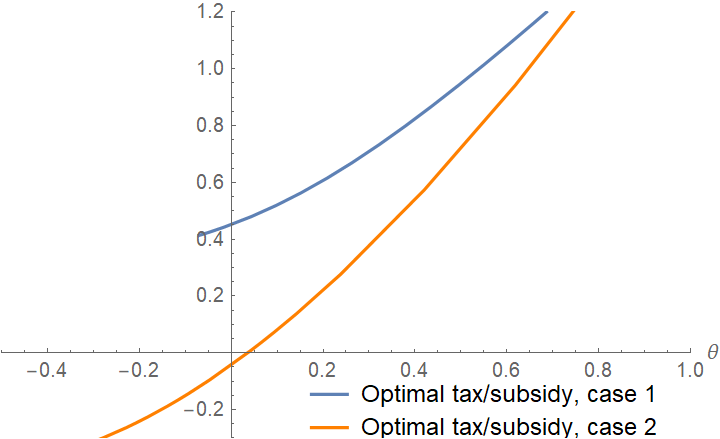}
    
    \caption{The optimal tax for different levels of overreaction $\theta$ and the two sets of parameters of Figures \ref{Fig:overunder}: case 1 corresponds to Figure \ref{Fig:overreaction}, case 2 corresponds to Figure \ref{Fig:underreaction}.
    In particular, the range of $\theta$ for case 1 is shorter because for $\theta$
 smaller than -0.1 the optimal tax would have been smaller than $-\beta$, and so not feasible.    }
    
    \label{fig:tax}
\end{figure}

Proposition \ref{prop_tax} addresses the problem of the introduction of a small tax. 
Figure \ref{fig:tax} represents instead the numerical calculation of the \emph{optimal} tax level (not infinitesimal) for 
 different values of $\theta$, and for the two sets of parameters of Figures \ref{Fig:overunder} (labeled here case 1 and case 2). We can see that, consistently with the intuition, the tax is increasing with $\theta$, and for low values (large enough underreaction), in case 2 the optimal tax is negative, i.e. a subsidy.

\subsubsection{Tax affecting only informed traders}

In this paragraph we explore a variation in which it is possible to levy the tax only on informed speculators, and we show that the qualitative results are very similar.

If the tax affects only the informed speculators, the liquidity suppliers inverse demand remains $p=-\mu_S-S+\beta\overline{D}$ as in the baseline model. Instead, the loadings in the informed traders strategies are given by expressions \eqref{demand_tax}, and the coefficient $a(\delta)$ solves the equation:
\[
(\gamma+\delta) a(\delta)=\frac{\tau_{\varepsilon}}{\tau_{\varepsilon}+\tau(a(\delta))}
\]
with the difference that now the precision of public information does not depend directly on $\delta$: $\tau(a(\delta))=a(\delta)^2(\theta+1)^2\beta^2\tau_S$. As a consequence, $\delta$ decreases both the loading on private and the loading on public information.

The results are collected in the following Proposition.

\begin{prop}

The welfare loss from the introduction of a tax $\delta$ is:
\begin{align*}
WL^{\delta}&=\frac{1}{2}\left(\frac{(1-(\gamma+\delta) \alpha)}{\left(\beta+\gamma+\delta\right)^2\tau}\left((1-(\gamma+\delta) \alpha)(\beta+\gamma)+2\delta(1+\beta\alpha)\right)+\frac{\delta^2(\mu_S^2+\tau_S^{-1}+\tau_0^{-1})}{(\beta+\gamma)(\beta+\gamma+\delta)^2}\right.\\
&\left.+\left(\frac{1}{\beta+\gamma+\delta}\right)^2\left(\theta^2(\beta+\gamma)-2\delta\theta\left(1-\frac{\tau_0}{\beta\alpha \tau_S}\right)\right)\left(\frac{1}{\tau_0}-\frac{1}{\tau}\right)+\frac{\gamma \alpha^2}{\tau_{\varepsilon}}\right)
\end{align*}

\begin{enumerate}

\item If $\theta$ is large enough, the introduction of a small tax is welfare improving.

\item If $\theta$ is small enough, the introduction of a small subsidy is welfare improving.

\item if $\theta=0$, the introduction of a small tax is welfare-improving if and only if $a^*>a^T$.

\end{enumerate}

\label{prop_tax2}
\end{prop}

\added{The only qualitative difference from Proposition \ref{prop_tax} is point 3, saying that the first order effect of the tax when $\theta=0$ is determined by whether the learning or the pecuniary externality dominates in the Bayesian benchmark. This is true in this case because the effect of the tax acts only through the loadings of the demand of the informed traders, and the loadings are all at the optimal level exactly when $a^*=a^T$.}

\section{Conclusion}
\label{conclusion}

We show that overreaction to information in the form of diagnostic expectations can improve welfare in markets where there is a strong enough information externality. When the information externality is not strong enough,
overreaction can rationalize a tax on financial transactions on efficiency grounds. These results highlight that understanding the \emph{degree} of overreaction is crucial for understanding its welfare effect and the sign of the optimal intervention.
The interactions of these effects with other rationales for trading, such as hedging or heterogeneity, and other biases such as cursedness, are interesting avenues for further research. 

\bibliographystyle{apalike}
\bibliography{biblio}

\appendix

\clearpage

\section*{Appendix}

\section{Alternative interpretation for the liquidity suppliers}
\label{liquidity}

In this section we illustrate an alternative interpretation for the origin of the elastic inverse demand, originating from a simple reduced form model of an entrepreneur issuing equity. There is an entrepreneur that has a project with dividend value $V$, that is not ex-ante known. The entrepreneur has preferences for remaining in control of the firm, measured by the random variable $\mu_S+S$, that represents the disutility per share sold for the entrepreneur. If she sells an amount $\overline{D}$ of equity, she can raise $p\overline{D}$, at the utility cost $(\mu_S+S)\overline{D}$, paying the transaction costs $\frac{\beta}{2}\overline{D}^2$. So, in total, the profit of the entrepreneur is:
\[
u_i^{e}=(p+\mu_S+S)\overline{D}-\frac{\beta}{2}\overline{D}^2
\]
that gives rise exactly to the inverse demand in the main text.

\newpage 

\section{Proofs}

\subsection{Proof of Proposition \ref{decentralized_solution1}}

We assume that $p=A+BV-CS$, and then derive the paramters such that this is an equilibrium.
The optimal choice for agents is:
	\[
D_i=\frac{1}{\gamma}\left(\frac{(\theta+1)\tau_{\varepsilon }}{\tau_{\varepsilon }+\tau_0+\tau_{p\mid V}}s_i+\frac{(\theta+1)\tau_{p\mid V }}{\tau_{\varepsilon }+\tau_0+\tau_{p\mid V}}(p-A)/B-p\right) 
\]
So the loadings are:
\[
\alpha=\frac{1}{\gamma}\frac{(\theta+1)\tau_{\varepsilon }}{\tau_{\varepsilon }+\tau_0+\tau_{p\mid V}}
\]
\[
\eta=\frac{(\theta+1)\tau_{p\mid V }}{\tau_{\varepsilon }+\tau_0+\tau_{p\mid V}}\dfrac{1}{B}
\]

Now, we solve for the equilibrium for generic coefficients $\alpha$ and $\eta$. This will be helpful in clarifying the intuitions later on.

If $p=A+BV-CS$ then $\tau_{p\mid V}=B^2/C^2\tau_S$. 

So the market clearing reads:
	\[
	p=-\mu_S-S+\beta/\gamma (\gamma \alpha V+\gamma \eta (p-A)/B-p)
	\]
 Solving for $p$:
\[
p=\frac{-\gamma \mu_S-\gamma S+\beta (\gamma \alpha V-\gamma \eta A/B)}{\gamma+\beta(1-\gamma\eta/B) }
\]	
So:
\[
B=\frac{\beta \gamma \alpha}{\gamma+\beta(1-\gamma\eta/B) }
\]
\[
1=\frac{\beta \gamma \alpha }{B(\gamma+\beta)-\beta\gamma\eta }
\]
\[
B=\beta\gamma\frac{\alpha+\eta}{\gamma+\beta}
\]
\[
C=\frac{\gamma}{\gamma+\beta(1-\gamma \eta /B) }
\]
\[
C=\frac{\gamma B}{(\gamma+\beta)B-\gamma\beta\eta}=\frac{\gamma \beta\gamma\frac{\alpha+\eta}{\gamma+\beta}}{(\gamma+\beta)\beta\gamma\frac{\alpha+\eta}{\gamma+\beta}-\gamma\beta\eta}=\frac{\gamma (\alpha+\eta)}{(\gamma+\beta)\alpha}
\]
so that: $B^2/C^2=\beta^2\alpha^2$, and so: $\tau_{p\mid V}=\alpha^2\beta^2\tau_S$.

Now, in equilibrium, $a$ must satisfy:
\[
\gamma a=\frac{\tau_{\varepsilon }}{\tau_{\varepsilon }+\tau_0+B^2/C^2\tau_S}=\frac{\tau_{\varepsilon }}{\tau_{\varepsilon }+\tau_0+\beta^2a^2(\theta+1)^2\tau_S}
\]
Define $\tau(a)=\beta^2a^2(\theta+1)^2\tau_S$ the precision of public information. Since the RHS is monotone decreasing and the LHS is monotone increasing (from 0 to $\infty$), there is a unique positive solution. 

The loading on public information is $\alpha=a(\theta+1)$, while the loading on public information is: $\eta=(1-\gamma \alpha)/\gamma$. Using this relation, we get the equilibrium coefficients:

\[
B=\beta\frac{(\theta+1))}{\gamma+\beta}
\]
\[
C=
\frac{\theta+1}{\alpha(\gamma+\beta)}
\]
\[
A=\frac{-\gamma\mu_S-\beta((1-\gamma a)(\theta+1) A/B)}{\gamma+\beta(1-(1-\gamma a)(\theta+1)/B) }
\]
\[
A+\frac{\beta ((1-\gamma a)(\theta+1) A/B)}{\gamma+\beta(1-(1-\gamma a)(\theta+1)/B) }=\frac{-\gamma \mu_S}{\gamma+\beta(1-(1-\gamma a)(\theta+1)/B) }
\]
\[
A=\frac{-\gamma \mu_S}{\gamma+\beta}
\]

Using the Law of the Large Numbers, we can express the total demand as:
\[
\overline{D}=\frac{\gamma \alpha V+\gamma \eta\EE(V\mid p)+S+\mu_S}{\beta+\gamma}
\]
or:
\[
\overline{D}=\frac{\gamma a(\theta+1) V+(1-\gamma a)(\theta+1)\EE(V\mid p)+S+\mu_S}{\beta+\gamma}
\]

\qed

\subsection{Proof of Corollary \ref{positive}}

\begin{enumerate}
    
    \item The first point follows from the implicit function theorem. Indeed, we have: 
    \[
\frac{\dd a }{\dd \theta}=-\frac{\frac{2 a^2 \beta ^2
   (\theta +1) \tau _S \tau
   _{\epsilon }}{\left(a^2
   \beta ^2 (\theta +1)^2 \tau
   _S+\tau _0+\tau _{\epsilon
   }\right){}^2}}{\frac{2 a \beta ^2 (\theta
   +1)^2 \tau _S \tau
   _{\epsilon }}{\left(a^2
   \beta ^2 (\theta +1)^2 \tau
   _S+\tau _0+\tau _{\epsilon
   }\right){}^2}+\gamma
}=-\frac{2\gamma a (1-\gamma a)\frac{\tau-\tau_0}{\tau(\theta+1)}}{2\gamma (1-\gamma a)\frac{\tau-\tau_0}{\tau}+\gamma }=-\frac{2 a (1-\gamma a)\frac{\tau-\tau_0}{\tau(\theta+1)}}{2 (1-\gamma a)\frac{\tau-\tau_0}{\tau}+1 }
\]
so $\frac{\dd a }{\dd \theta}<0$. But:
\[
\frac{\dd \alpha }{\dd \theta}=\frac{\dd a }{\dd \theta}(\theta+1)+a=-a\frac{2\gamma  (1-\gamma a)\frac{\tau-\tau_0}{\tau}}{2\gamma (1-\gamma a)\frac{\tau-\tau_0}{\tau}+\gamma }+a>0
\]
    
    \item from the proof of Proposition \ref{decentralized_solution1} we get that $B^2/C^2=a^2(\theta+1)^2\beta^2\tau_S=\alpha^2\beta^2\tau_S$, hence it is increasing in $\theta$.

\item the volatility of the price is given by:
\[
Var(p)=B^2+C^2=\frac{1}{(\gamma+\beta)^2}\left(\beta^2(\theta+1)^2\frac{1}{\tau_0}+\frac{1}{a^2\tau_S}\right)
\]
that is increasing in $\theta$.
    
\end{enumerate}
 
\qed 

\subsection{Proof of Proposition \ref{bayesian}}

First, we compute the limits of $\alpha$ at the extreme of the domain.
For $\theta\to -1$ we have that $a$ goes to its maximum, $\underline{a}=\frac{\tau_{\varepsilon }}{\tau_{\varepsilon}+\tau_0}$, and $\alpha\to 0$, as $\tau$. For $\theta \to \infty$ instead we have $a \to 0$ but $\alpha \to \infty$. Indeed, both $a$ and $\alpha$ are monotonic so they have a limit. Indeed, if $\lim_{\theta\to \infty}a=a'>0$ (possibly infinite) we would have: 
\[
\lim_{\theta\to \infty}a=\lim_{\theta\to \infty}\frac{\tau_{\varepsilon}}{\gamma (\tau_{\varepsilon}+\tau_0(a')^2\beta^2(\theta+1)^2 )}=0
\]  
and if $\lim_{\theta\to \infty}\alpha=\alpha'<\infty$ (possibly zero), we would have:
\[
\lim_{\theta\to \infty}\alpha=\lim_{\theta\to \infty}\frac{\tau_{\varepsilon}(\theta+1)}{\gamma (\tau_{\varepsilon}+\tau_0(\alpha')^2\beta^2)}=\infty 
\]  
that would be contradictions. 

Now, for part 1, 
 the limits computed above show that it increases from 0 to infinity, so there is at least a value $\theta'$ satisfying the condition. Moreover, Corollary \ref{positive} shows that $\alpha$ is monotonic in $\theta$, so there can be only one.

 For part 2, the reasoning is analogous: the derivative of the loading is:
 \[
 \frac{\dd }{\dd \theta}\left(\dfrac{\theta+1}{\gamma}-\alpha\right)=\dfrac{1}{\gamma} -\frac{\dd \alpha }{\dd \theta}=\dfrac{1}{\gamma}-a+a\frac{2\gamma  (1-\gamma a)\frac{\tau-\tau_0}{\tau}}{2\gamma (1-\gamma a)\frac{\tau-\tau_0}{\tau}+\gamma }>\dfrac{1-\gamma a}{\gamma}>0
 \]
 so it is monotonically increasing. Moreover, for $\theta\to -1$ the loading goes to zero. Instead, for $\theta\to \infty$ we have that $\alpha\to \infty$, so:
 \[
\dfrac{\theta+1}{\gamma}-\alpha=\dfrac{\theta+1}{\gamma}\left(1- \frac{\tau_{\varepsilon}}{\tau_{\varepsilon}+\tau_0(\alpha)^2\beta^2}\right)
 \]
 the term in the parenthesis goes to 1 as $\alpha\to \infty$, so the loading diverges. So, the equation has one and only one solution.

For part 3, we have that:
\[
\dfrac{1}{\gamma}-\alpha(\theta')=\dfrac{\theta''+1}{\gamma}-\alpha(\theta'') \iff 
\]
\[
\alpha(\theta'')-\alpha(\theta')=\dfrac{\theta''}{\gamma}
\]
from which the thesis follows.

\qed

\subsection{Proof of Lemma \ref{decomposition}}

The expression for the welfare loss is:
\[
WL=W^*-W=(\beta+\gamma)\frac{1}{2}\EE(D^o-\overline{D})^2+\frac{\gamma}{2}\EE Var(D_i)
\]

The second term is:
\[
\EE(Var D_i)=\EE \int (-\alpha s_i+\alpha V)^2=\frac{\alpha^2}{\tau_{\varepsilon}}
\]

The first is:
\[
D^o-\overline{D}=\frac{V+\mu_S+S}{\beta+\gamma}-\frac{1}{\beta+\gamma}\left(\mu_S + S+\gamma \alpha V+\gamma \eta \EE(V\mid p)\right)
\]
\[
=\frac{(1-\gamma \alpha)}{\beta+\gamma}(V-\EE(V\mid p)))+\frac{1-\gamma \alpha-\gamma \eta}{\beta+\gamma}\EE(V\mid p))
\]
Now we want to compute the expectation of the square. This is equivalent to the variance since all the variables involved have zero expectation:
\begin{align*}
\EE(D^o-\overline{D})^2&=\frac{(1-\gamma a)^2}{(\beta+\gamma)^2}\EE(V-\EE(V\mid p))^2+\frac{(1-\gamma \alpha-\gamma \eta)^2}{(\beta+\gamma)^2}\EE(\EE(V\mid p)))^2\\
&+2\frac{(1-\gamma \alpha)(1-\gamma \alpha-\gamma \eta)}{(\beta+\gamma)^2}Cov((V-\EE(V\mid p))\EE(V\mid p))
\end{align*}
We are going to use the following facts:
\[
\EE(V-\EE(V\mid p))^2=\EE(\EE((V-\EE(V\mid p))^2\mid p))=\EE(Var(V\mid p))=\EE\left(\frac{1}{\tau}\right)=\frac{1}{\tau}
\]
\[
\EE(\EE(V\mid p)^2)=\frac{(\tau-\tau_0)^2}{\tau^2}\EE\left(V-\frac{C}{B}S\right)^2=\frac{(\tau-\tau_0)^2}{\tau^2}\left(\frac{1}{\tau_0}+\frac{C^2}{B^2}\frac{1}{\tau_S}\right)=\frac{1}{\tau_0}-\frac{1}{\tau}
\]
and:
\[
Cov((V-\EE(V\mid p))\EE(V\mid p))=\EE(\EE(V\mid p)V)-\EE(\EE(V\mid p)^2)=0
\]

So:
\[
\EE(D^o-\overline{D})^2=\frac{(1-\gamma a)^2}{(\beta+\gamma)^2\tau}+\frac{(1-\gamma \alpha-\gamma \eta)^2}{(\beta+\gamma)^2}\left(\frac{1}{\tau_0}-\frac{1}{\tau}\right)
\]

So, the total welfare loss is:
\[
WL=\frac{1}{2}\frac{(1-\gamma \alpha)^2}{(\beta+\gamma)}\frac{1}{\tau}+\frac{1}{2}\frac{(1-\gamma \alpha-\gamma \eta)^2}{(\beta+\gamma)}\left(\frac{1}{\tau_0}-\frac{1}{\tau}\right)+\frac{\gamma \alpha^2}{2\tau_{\varepsilon}}
\]
where from Proposition \ref{decentralized_solution1}, we have that $1-\gamma \alpha-\gamma \eta=\theta$. 
So it can be decomposed as:
\[
WL^B(\alpha)=\frac{1}{2}\frac{(1-\gamma \alpha)^2}{(\beta+\gamma)}\frac{1}{\tau}+\frac{\gamma \alpha^2}{2\tau_{\varepsilon}}
\]
\[
WL^{D}=\frac{1}{2}\frac{(1-\gamma \alpha-\gamma \eta)^2}{(\beta+\gamma)}\left(\frac{1}{\tau_0}-\frac{1}{\tau}\right)=\frac{1}{2}\frac{\theta^2}{(\beta+\gamma)}\left(\frac{1}{\tau_0}-\frac{1}{\tau}\right)
\]

\qed

\subsection{Proof of Proposition \ref{welfare}}

From Lemma \ref{decomposition}, we have that the welfare loss has two components:
\[
WL=WL^B+WL^D
\]
where $WL^B$ depends on $\theta$ only via $\alpha$, and $WL^D$ is second order in $\theta$. Hence, in $\theta=0$:
\[
\frac{\dd WL}{\dd \theta}\mid_{\theta=0}=\frac{\partial WL^B}{\partial \alpha}\frac{\dd \alpha}{\dd \theta}\mid_{\theta=0}
\]
Moreover, from Corollary \ref{positive} we know that $\alpha$ is increasing in $\theta$, so we conclude that, in $\theta=0$, $\frac{\dd WL}{\dd \theta}$ this has the same sign as $\frac{\partial WL^B}{\partial \alpha}$. Since this is the Bayesian welfare loss, this is positive if and only if $a^*>a^T$.

The derivatives are:
\[
\frac{\partial WL^B}{\partial \alpha}=-\frac{
   (1-\alpha  \gamma)\left(\gamma  \tau _0
   +\alpha 
   \beta ^2 \tau _S\right)}{(\beta
   +\gamma ) \left(\alpha ^2
   \beta ^2 \tau _S+\tau
   _0\right){}^2}+\frac{ \alpha  \gamma }{\tau
   _{\epsilon }}
\]
\[
\frac{\partial WL^D}{\partial \theta}=\added{ \frac{\theta}{\beta+\gamma}\left(\frac{1}{\tau_0}-\frac{1}{\tau}\right) }
\]

\[
\frac{\partial WL^D}{\partial \alpha}=\added{\frac{ \alpha  \beta ^2
   \theta ^2 \tau _S}{(\beta
   +\gamma ) \left(\alpha ^2
   \beta ^2 \tau _S+\tau
   _0\right){}^2}}
\]

Using the limits computed in the proof of Proposition \ref{bayesian}, it follows that for $\theta\to -1$ $\frac{\dd \alpha}{\dd \theta}$ goes to the finite value $\underline{a}>0$, while for $\theta\to \infty$ it goes to zero.

\added{Now for $\theta\to \infty$ we have that $\frac{\partial WL^D}{\partial \theta}$ goes to infinity for $\theta\to \infty$, $\frac{\partial WL^D}{\partial \alpha}>0$ and $\frac{\partial WL^B}{\partial \alpha}>0$ because $WL^B$ has a finite minimum. So we conclude that $\lim_{\theta\to \infty}\frac{\dd WL}{\dd \theta}=+\infty$.}

\added{Instead, for $\theta\to -1$ $\frac{\partial WL^D}{\partial \theta}<0$, $\frac{\partial WL^D}{\partial \alpha}$ goes to zero, and $\frac{\dd \alpha}{\dd \theta}$ goes to the finite value $\underline{a}>0$. So only $\frac{\partial WL^D}{\partial \theta}<0$ survives, and the limit is negative: $\lim_{\theta\to -1}\frac{\dd WL^D}{\dd \theta}<0$}

Now for $\theta\to \infty$ the welfare loss diverges: hence the optimal value of $\theta$ has to be finite. (take any finite value $t=WL(\theta')$, there is a $\theta''$ such that $WL>t$ for all $\theta>\theta''$ and so the optimum is smaller than $\theta''$). Hence, for $\theta$ large enough, $\frac{\dd WL}{\dd \theta}>0$.

\qed

\subsection{Proof of Proposition \ref{loading_tax}}

All the equilibrium expressions are analogous to what derived in Proposition \ref{decentralized_solution1}, with $\gamma+\delta$ in place of $\gamma$ and $\beta+\delta$ in place of $\beta$.

The level of trade for agent $i$ is:
\[
D_i=(\theta+1)a s_i+\frac{(1-(\gamma+\delta) a)(\theta+1)\EE(V\mid p)-p}{\gamma+\delta}
\]
where $a$ solves:
\[
(\gamma+\delta) a=\frac{\tau_{\varepsilon}}{\tau_{\varepsilon}+\tau(a)}
\]
and $\tau(a)=a^2(\theta+1)^2(\beta+\delta)^2\tau_S$.

Using the implicit function theorem, the effect of $\delta$ on the loading on private information is:
\[
\frac{\dd \alpha}{\dd \delta}=\added{-\frac{\alpha  \left(\alpha ^2
   (\beta +\delta ) \tau _S
   (\beta +2 (\gamma +\delta
   )+\delta )+\tau _0+\tau
   _{\varepsilon
   }\right)}{(\gamma +\delta )
   \left(3 \alpha ^2 (\beta
   +\delta )^2 \tau _S+\tau
   _0+\tau _{\varepsilon
   }\right)}}<0
\]
So it is monotonic in $\delta$. Moreover, the limit of $\alpha$ for $\delta\to \infty$ is zero, and the limit for $\delta\to-\gamma$ is $+\infty$ (the proof of these two statements is below). So, there always is a unique $\delta^*$ such that $\alpha(\delta^*)=a^T$, proving point 1.

If the limit for $\delta\to\infty$ was a finite or infinite value $\alpha'>0$ we would have:
\[
\lim_{\delta\to \infty} \alpha=\lim_{\delta\to \infty}\dfrac{1}{\gamma+\delta}\dfrac{(\theta+1)\tau_{\varepsilon}}{\tau_{\varepsilon}+\tau(\alpha'))}=\dfrac{(\theta+1)\tau_{\varepsilon}}{\tau_{\varepsilon}+\tau(\alpha'))}\lim_{\delta\to \infty}\dfrac{1}{\gamma+\delta}=0=\alpha' 
\]
that would contradict $\alpha' >0$.

The limit for $\delta\to-\gamma$ is $+\infty$. Indeed, if it was a finite value $\alpha'$, as above:
\[
\lim_{\delta\to -\gamma} \alpha=\lim_{\delta\to -\gamma}\dfrac{1}{\gamma+\delta}\dfrac{(\theta+1)\tau_{\varepsilon}}{\tau_{\varepsilon}+\tau(\alpha'))}=\dfrac{(\theta+1)\tau_{\varepsilon}}{\tau_{\varepsilon}+\tau(\alpha'))}\lim_{\delta\to -\gamma}\dfrac{1}{\gamma+\delta}=+\infty=\alpha'
\]
that would contradict the fact that $\alpha'$ is finite.

The effect of the tax on the loading on public information is, instead:
\begin{align*}
\frac{\dd }{\dd \delta}\left(\dfrac{\theta+1}{\gamma+\delta}-\alpha\right)&=-\dfrac{\theta+1}{(\gamma+\delta)^2}+\frac{\alpha  \left(\alpha ^2
   (\beta +\delta ) \tau _S
   (\beta +2 (\gamma +\delta
   )+\delta )+\tau _0+\tau
   _{\varepsilon
   }\right)}{(\gamma +\delta )
   \left(3 \alpha ^2 (\beta
   +\delta )^2 \tau _S+\tau
   _0+\tau _{\varepsilon
   }\right)}    \\
   &=-\dfrac{\theta+1}{(\gamma+\delta)^2}\left(1-\dfrac{\tau_{\varepsilon}}{\tau_{\varepsilon}+\tau}\frac{  \left(\alpha ^2
   (\beta +\delta ) \tau _S
   (\beta +2 (\gamma +\delta
   )+\delta )+\tau _0+\tau
   _{\varepsilon
   }\right)}{
   \left(3 \alpha ^2 (\beta
   +\delta )^2 \tau _S+\tau
   _0+\tau _{\varepsilon
   }\right)}\right)\\
     &=-\dfrac{\theta+1}{(\gamma+\delta)^2}\left(\dfrac{\tau}{\tau_{\varepsilon}+\tau}-\frac{  2\alpha ^2
   (\beta +\delta ) \tau _S
   (\gamma -\beta)
}{
   \left(3 \alpha ^2 (\beta
   +\delta )^2 \tau _S+\tau
   _0+\tau _{\varepsilon
   }\right)}\right)
\end{align*}
If $\gamma<\beta$ we have that the derivative is negative. Instead, for $\gamma>\beta$, we have that the derivative is positive if and only if:
\[
\dfrac{\tau}{\tau_{\varepsilon}+\tau}-\frac{  2\alpha ^2
   (\beta +\delta ) \tau _S
   (\gamma -\beta)
}{
   \left(3 \alpha ^2 (\beta
   +\delta )^2 \tau _S+\tau
   _0+\tau _{\varepsilon
   }\right)}\le \dfrac{\tau}{\tau_{\varepsilon}+\tau}-\frac{  2(\tau-\tau_0))
   (\gamma -\beta)
}{3(\tau+\tau_{\varepsilon})}
\]
that is negative if and only if $\gamma>\beta+\dfrac{3\tau}{2(\tau-\tau_0)}$. The LHS grows from zero to $\infty$ and the RHS decreases from $\infty$ to zero, so it follows that for $\gamma$ large enough this is satisfied, proving point 2.

Part 3 and 4 are immediate from $\eta_p=\dfrac{1}{\gamma+\delta}$

\qed

\subsection{Proof of Proposition \ref{prop_tax}}

All the equilibrium expressions are analogous to what derived in Proposition \ref{decentralized_solution1}, with $\gamma+\delta$ in place of $\gamma$ and $\beta+\delta$ in place of $\beta$.

The level of trade for agent $i$ is:
\[
D_i=(\theta+1)a s_i+\frac{(1-(\gamma+\delta) a)(\theta+1)\EE(V\mid p)-p}{\gamma+\delta}
\]
where $a$ solves:
\[
(\gamma+\delta) a=\frac{\tau_{\varepsilon}}{\tau_{\varepsilon}+\tau(a)}
\]
and $\tau(a)=a^2(\theta+1)^2(\beta+\delta)^2\tau_S$, and:
\[
\overline{D}=\frac{1}{\beta+\gamma+2\delta}\left(S+\mu_S+(\gamma+\delta)a (\theta+1)V+(1-(\gamma+\delta)a)(\theta+1)\EE(V\mid p)\right)
\]

From Lemma \ref{welfare_loss_vives}, we know that the expression for the welfare loss is:
\[
\frac{1}{2}\left(	(\beta+\gamma)\EE(D^o-\overline{D})^2 + \gamma \EE Var(D_i) \right)
\]
this is not affected, because the lump-sum rebate means that the tax terms cancel out.

The first best solution $D^o$ is of course not affected by the tax. We have to compute the two terms using the individual demands under a tax $\delta$. The dispersion term has the same form as a function of $a$ as would without the tax:
\begin{align*}
    \EE Var(D_i)&=\EE \int \alpha^2 (s_i-V)^2\dd i=\alpha^2\int \EE(s_i-V)^2\dd i\\
    &=\alpha^2\int \EE(\EE((s_i-V)^2\mid V))\dd i=\frac{\alpha^2}{\tau_{\varepsilon}}
\end{align*}

Instead, for the volatility term:
\begin{align*}
\overline{D}^o-\overline{D}&=\frac{\mu_S+S+V}{\beta+\gamma}-\frac{1}{\beta+\gamma+2\delta}\left(S+\mu_S+(\gamma+\delta)a (\theta+1)V+(1-(\gamma+\delta)a)(\theta+1)\EE(V\mid p)\right) \\
&
=\frac{(\beta+\gamma)((1-(\gamma+\delta)a (\theta+1)V)+(1-(\gamma+\delta)a)(\theta+1)\EE(V\mid p))+2\delta(\mu_S+S+V)}{(\beta+\gamma)(\beta+\gamma+2\delta)}\\
&
=\frac{1}{(\beta+\gamma)(\beta+\gamma+2\delta)}\left((\beta+\gamma)(1-(\gamma+\delta)\alpha) (V-\EE(V\mid p))+\right. \\
&
\left. +(\beta+\gamma)\theta\EE(V\mid p)+2\delta(\mu_S+V+S)\right)
\end{align*}

Taking the square and the expectation we get:
\begin{align*}
\EE(D^o-\overline{D})^2&=\frac{(1-(\gamma+\delta) \alpha)}{\left(\beta+\gamma+2\delta\right)^2\tau}\left((1-(\gamma+\delta) \alpha)+\frac{4\delta}{\beta+\gamma}(1+(\beta+\delta)\alpha)\right)+\frac{4\delta^2(\mu_S^2+\tau_S^{-1}+\tau_0^{-1})}{(\beta+\gamma)^2(\beta+\gamma+2\delta)^2}\\
&+\left(\frac{1}{\beta+\gamma+2\delta}\right)^2\left(\theta^2-\frac{4\delta\theta}{\beta+\gamma}\left(1-\frac{\tau_0}{(\beta+\delta)\alpha \tau_S}\right)\right)\left(\frac{1}{\tau_0}-\frac{1}{\tau}\right)
\end{align*}
So the total welfare loss is:
\begin{align*}
WL^{\delta}&=\dfrac{1}{2}\left(\frac{(1-(\gamma+\delta) \alpha)}{\left(\beta+\gamma+2\delta\right)^2\tau}\left((1-(\gamma+\delta) \alpha)(\beta+\gamma)+4\delta(1+(\beta+\delta)\alpha)\right)+\frac{4\delta^2(\mu_S^2+\tau_S^{-1}+\tau_0^{-1})}{(\beta+\gamma)(\beta+\gamma+2\delta)^2}\right.\\
&\left.+\left(\frac{1}{\beta+\gamma+2\delta}\right)^2\left(\theta^2(\beta+\gamma)-4\delta\theta\left(1-\frac{\tau_0}{(\beta+\delta)\alpha \tau_S}\right)\right)\left(\frac{1}{\tau_0}-\frac{1}{\tau}\right)+\frac{\gamma \alpha^2}{\tau_{\varepsilon}}\right)
\end{align*}


Calculating the derivatives in $\delta=0$ we get:
\begin{align}
 \frac{\partial WL^{\delta}}{\partial \alpha}   &=-\frac{
   (1-\alpha  \gamma)\left(\gamma  \tau _0
   +\alpha 
   \beta ^2 \tau _S\right)+\theta^2\alpha 
   \beta ^2 \tau _S}{(\beta
   +\gamma ) \left(\alpha ^2
   \beta ^2 \tau _S+\tau
   _0\right){}^2}+\frac{ \alpha  \gamma }{\tau
   _{\epsilon }}
 \nonumber \\ 
 \frac{\partial WL^{\delta}}{\partial \delta}    &= \frac{ \alpha  \left(\tau
   _0^2 (1-(\alpha  \gamma )
   (\beta +\gamma )+2 \beta 
   \theta )-2 \alpha ^3 \beta
   ^4 \theta  (\theta +1) \tau
   _S^2\right.}{\tau _0
   (\beta +\gamma )^2
   \left(\alpha ^2 \beta ^2
   \tau _S+\tau _0\right){}^2} \nonumber\\
   &+\frac{\left.-\alpha  \beta  \tau _0
   \tau _S \left((\alpha 
   \gamma -1) (\beta +\gamma )
   (\alpha  (\beta +\gamma
   )-1)-2 \beta  \theta 
   (\alpha  \beta -1)+\theta
   ^2 (\beta -\gamma
   )\right)\right)}{\tau _0
   (\beta +\gamma )^2
   \left(\alpha ^2 \beta ^2
   \tau _S+\tau _0\right){}^2}
   \label{deltazero2}
\end{align}

Now consider part 1 of the result.
\added{In the limit for $\theta\to \infty$ we have that $\frac{\dd \alpha}{\dd \delta}$ diverges negatively, while in $ \frac{\partial WL^{\delta}}{\partial \delta}$ the leading term is:
\[
-\dfrac{2 \alpha^4\beta ^4 \theta  (\theta
   +1) }{\alpha^4\beta ^4\tau_S^4}\to -\infty
\]
Moreover, we have seen in Proposition \ref{welfare} that $\frac{\partial WL^{\delta}}{\partial \alpha}>0$ for $\theta$ small enough: so we get that $\frac{\partial WL^{\delta}}{\partial \delta}<0$ for $\theta$ large enough.}

Consider part 2.
The total derivative goes to zero as $\theta\to -1$ (and $\alpha\to 0$). We can observe that both $\frac{\partial WL}{\partial \delta}$ and $\frac{\dd \alpha}{\dd \delta}$ have a factor of $\alpha$.  So, we collect $\alpha$, and calculating we get:
\[
\lim_{\theta\to -1}\frac{1}{\alpha}\frac{\dd WL^{\delta}}{\dd \delta}=\lim_{\theta\to -1}\frac{1}{\alpha}\left(\frac{\partial WL^{\delta}}{\partial \delta}+\frac{\partial WL^{\delta}}{\partial \alpha}\frac{\dd \alpha}{\dd \delta}\right)=\replaced{\dfrac{4\gamma}{(\beta+\gamma)^2\tau_0}}{\frac{2\left(\tau _0 (\beta +\gamma
   )+\beta \right)}{\tau _0^2 (\beta
   +\gamma )^2}}>0
\]

Now consider part 3. If $\theta=0$ expression \ref{deltazero2} shows that $\frac{\partial WL^{\delta}}{\partial \delta}=\frac{ a^*  (1- 
   \gamma a^*) \left(a^* 
   \beta  \tau _S (a^* 
   (\beta +\gamma )-1)+\tau
   _0\right)}{(\beta +\gamma )
   \left((a^*) ^2 \beta ^2
   \tau _S+\tau _0\right){}^2}$. If $a^*=a^T$, by definition, $\frac{\partial WL^{\delta}}{\partial \alpha}=0$, and $\frac{\dd \alpha}{\dd \delta}$ remains finite. So the total derivative is positive \replaced{if and only if $\alpha 
   \beta  \tau _S (\alpha 
   (\beta +\gamma )-1)+\tau
   _0>0$}{$\frac{\dd WL}{\dd \delta}>0$}.

\qed 

\subsection{Proof of Proposition \ref{prop_tax2}}
    
All the equilibrium expressions are analogous to what derived in Proposition \ref{decentralized_solution1}, with $\gamma+\delta$ in place of $\gamma$.

The level of trade for agent $i$ is:
\[
D_i=(\theta+1)a s_i+\frac{(1-(\gamma+\delta) a)(\theta+1)\EE(V\mid p)-p}{\gamma+\delta}
\]
where $a$ solves:
\[
(\gamma+\delta) a=\frac{\tau_{\varepsilon}}{\tau_{\varepsilon}+\tau(a)}
\]

From Lemma \ref{welfare_loss_vives}, we know that the expression for the welfare loss is:
\[
\frac{1}{2}\left(	(\beta+\gamma)\EE(D^o-\overline{D})^2 + \gamma \EE Var(D_i) \right)
\]

The first best solution $D^o$ is of course not affected by the tax. We have to compute the two terms using the individual demands under a tax $\delta$. The dispersion term has the same form as a function of $a$ as would without the tax.

Instead, for the volatility term:
\[
\overline{D}=\frac{1}{\beta+\gamma+\delta}\left(S+\mu_S+(\gamma+\delta)a (\theta+1)V+(1-(\gamma+\delta)a)(\theta+1)\EE(V\mid p)\right)
\]
\begin{align*}
\overline{D}^o-\overline{D}&=\frac{\mu_S+S+V}{\beta+\gamma}-\frac{1}{\beta+\gamma+\delta}\left(S+\mu_S+(\gamma+\delta)a (\theta+1)V+(1-(\gamma+\delta))(\theta+1)\EE(V\mid p)\right) \\
&
=\frac{(\beta+\gamma)((1-(\gamma+\delta)a (\theta+1)V)+(1-(\gamma+\delta)a)(\theta+1)\EE(V\mid p))+\delta(\mu_S+S+V)}{(\beta+\gamma)(\beta+\gamma+\delta)}\\
&
=\frac{1}{(\beta+\gamma)(\beta+\gamma+\delta)}\left((\beta+\gamma)(1-(\gamma+\delta)\alpha) (V-\EE(V\mid p))+\right. \\
&
\left. +(\beta+\gamma)\theta\EE(V\mid p)+\delta(\mu_S+V+S)\right)
\end{align*}

Taking the square and the expectation we get:
\begin{align*}
\EE(D^o-\overline{D})^2&=\frac{(1-(\gamma+\delta) \alpha)}{\left(\beta+\gamma+\delta\right)^2\tau}\left((1-(\gamma+\delta) \alpha)+\frac{2\delta}{\beta+\gamma}(1+\beta\alpha)\right)+\frac{\delta^2(\mu_S^2+\tau_S^{-1}+\tau_0^{-1})}{(\beta+\gamma)^2(\beta+\gamma+\delta)^2}\\
&+\left(\frac{1}{\beta+\gamma+\delta}\right)^2\left(\theta^2-\frac{2\delta\theta}{\beta+\gamma}\left(1-\frac{\tau_0}{\beta\alpha \tau_S}\right)\right)\left(\frac{1}{\tau_0}-\frac{1}{\tau}\right)
\end{align*}
So the total welfare loss is:
\begin{align*}
WL^{\delta}&=\frac{1}{2}\left(\frac{(1-(\gamma+\delta) \alpha)}{\left(\beta+\gamma+\delta\right)^2\tau}\left((1-(\gamma+\delta) \alpha)(\beta+\gamma)+2\delta(1+\beta\alpha)\right)+\frac{\delta^2(\mu_S^2+\tau_S^{-1}+\tau_0^{-1})}{(\beta+\gamma)(\beta+\gamma+\delta)^2}\right.\\
&\left.+\left(\frac{1}{\beta+\gamma+\delta}\right)^2\left(\theta^2(\beta+\gamma)-2\delta\theta\left(1-\frac{\tau_0}{\beta\alpha \tau_S}\right)\right)\left(\frac{1}{\tau_0}-\frac{1}{\tau}\right)+\frac{\gamma \alpha^2}{\tau_{\varepsilon}}\right)
\end{align*}


Using the implicit function theorem, the effect of $\delta$ on the loadings is:
\[
\frac{\dd \alpha}{\dd \delta}=-\frac{\alpha}{(\gamma+\delta)  \left(\frac{2
   \alpha ^2 \beta ^2 (\theta
   +1)^2 \tau _S}{\alpha ^2
   \beta ^2 (\theta +1)^2 \tau
   _S+\tau _0+\tau _{\epsilon
   }}+1\right)}<0
\]
\begin{align*}
\frac{\dd \eta}{\dd \delta}=&-\dfrac{\theta+1}{(\gamma+\delta)^2}+\frac{\alpha}{(\gamma+\delta)  \left(\frac{2
   \alpha ^2 \beta ^2 (\theta
   +1)^2 \tau _S}{\alpha ^2
   \beta ^2 (\theta +1)^2 \tau
   _S+\tau _0+\tau _{\epsilon
   }}+1\right)}\\
   =&-\dfrac{\theta+1}{(\gamma+\delta)^2}\left(1-\dfrac{\tau_{\varepsilon}}{\tau+\tau_{\varepsilon}}\frac{1}{  \left(\frac{2
   \alpha ^2 \beta ^2 (\theta
   +1)^2 \tau _S}{\alpha ^2
   \beta ^2 (\theta +1)^2 \tau
   _S+\tau _0+\tau _{\epsilon
   }}+1\right)}\right)<0
\end{align*}

For $\theta\to \infty$ we can see that since $\alpha\to \infty$ we have $\frac{\dd \alpha}{\dd \delta}\to-\infty$.
For $\theta\to -1$ since $\alpha\to 0$ we have $\frac{\dd \alpha}{\dd \delta}\to 0$.

Calculating the derivatives in $\delta=0$ we get:
\begin{align}
   \frac{\partial WL^{\delta}}{\partial \alpha}\mid_{\delta=0}&= -\frac{
   (1-\alpha  \gamma)\left(\gamma  \tau _0
   +\alpha 
   \beta ^2 \tau _S\right)+\theta^2\alpha 
   \beta ^2 \tau _S}{(\beta
   +\gamma ) \left(\alpha ^2
   \beta ^2 \tau _S+\tau
   _0\right){}^2}+\frac{ \alpha  \gamma }{\tau
   _{\epsilon }}\\
   \frac{\partial WL}{\partial \delta}\mid_{\delta=0}&=-\frac{\alpha  \beta 
   \theta  \left(\alpha  \beta  (\theta
   +1) \tau _S-\tau_0\right)}{\tau _0 (\beta
   +\gamma )^2 \left(\alpha ^2 \beta ^2
   \tau _S+\tau _0\right)}
\label{deltazero}
\end{align}

Now consider part 1 of the result. The total derivative is:
\[
 \frac{\dd WL^{\delta}}{\dd \delta}\mid_{\delta=0}= \frac{\partial WL^{\delta}}{\partial \delta}\mid_{\delta=0}+ \frac{\partial WL^{\delta}}{\partial \alpha}\frac{\dd \alpha}{\dd \delta}\mid_{\delta=0}
\]
We have $\frac{\partial WL^{\delta}}{\partial \alpha}\mid_{\delta=0}>0$, $\frac{\dd \alpha}{\dd \delta}\mid_{\delta=0}to -\infty$, and $\frac{\partial WL^{\delta}}{\partial \delta}\mid_{\delta=0}$ also goes to $-\infty$. So, the welfare loss is decreasing for $\theta$ high enough.

Consider part 2.
The total derivative goes to zero as $\theta\to -1$ (and $\alpha\to 0$). We can observe that both $\frac{\partial WL^{\delta}}{\partial \delta}\mid_{\delta=0}$ and $\frac{\dd \alpha}{\dd \delta}\mid_{\delta=0}$ have a factor of $\alpha$.  So, we collect $\alpha$, and calculating we get:
\[
\lim_{\theta\to -1}\frac{1}{\alpha}\frac{\dd WL^{\delta}}{\dd \delta}\mid_{\delta=0}=\lim_{\theta\to -1}\frac{1}{\alpha}\left(\frac{\partial WL^{\delta}}{\partial \delta}\mid_{\delta=0}+\frac{\partial WL^{\delta}}{\partial \alpha}\frac{\dd \alpha}{\dd \delta}\mid_{\delta=0}\right)=\frac{2\gamma
   }{\tau _0^2 (\beta +\gamma )^2}>0
\]

Now consider part 3. If $\theta=0$ expression \ref{deltazero} shows that $ \frac{\partial WL^{\delta}}{\partial \delta}\mid_{\theta=\delta=0}=0$. Moreover, for $\theta=0$ the welfare loss is the same function of $\alpha$ as the Bayesian, and we know from \cite{vives2017endogenous} that it is convex, with a minimum in $a^*=a^T$.

\qed

\end{document}